\documentclass[aps,prl,amsmath,twocolumn,superscriptaddress]{revtex4-1}
\usepackage{graphicx}
\usepackage{rotating}
\usepackage{dcolumn}
\usepackage{bm}
\usepackage{color}
\usepackage{mathptmx, textcomp}
\usepackage[latin1]{inputenc}
\usepackage{braket}
\usepackage{hyperref}

\graphicspath{ {images/} }

\begin{document}

\author{M. J. Mark}\affiliation{Institut f\"ur Experimentalphysik und Zentrum f\"ur Quantenphysik, Universit\"at Innsbruck, 6020 Innsbruck, Austria}\affiliation{Institut f\"ur Quantenoptik und Quanteninformation, \"Osterreichische  Akademie  der  Wissenschaften, 6020 Innsbruck, Austria}
\author{S. Flannigan}\affiliation{Department of Physics \& SUPA, University of Strathclyde, Glasgow, G4 0NG, Scotland, U.K.}
\author{F. Meinert}\affiliation{5. Physikalisches Institut and Center for Integrated Quantum Science and Technology, Universit\"{a}t Stuttgart, Pfaffenwaldring 57, 70569 Stuttgart, Germany}
\author{K. Jag-Lauber}\affiliation{Institut f\"ur Experimentalphysik und Zentrum f\"ur Quantenphysik, Universit\"at Innsbruck, 6020 Innsbruck, Austria}
\author{J. P. D'Incao}
\affiliation{JILA, National Institute of Standards and Technology, and the University of Colorado, Department of Physics, Boulder, Colorado 80309, USA}
\author{A.~J. Daley}\affiliation{Department of Physics \& SUPA, University of Strathclyde, Glasgow, G4 0NG, Scotland, U.K.}
\author{H.-C. N\"agerl}\affiliation{Institut f\"ur Experimentalphysik und Zentrum f\"ur Quantenphysik, Universit\"at Innsbruck, 6020 Innsbruck, Austria}

\title{Interplay between coherent and dissipative dynamics of bosonic doublons in an optical lattice}

\date{\today}

\begin{abstract}
We observe the dissipative dynamics of  a dense, strongly interacting gas of bosonic atom pairs in an optical lattice, controlling the strength of the two-body interactions over a wide parameter regime. We study how three-body losses contribute to the lattice dynamics, addressing a number of open questions related to the effects of strong dissipation in a many-body system, including the relationship to the continuous quantum Zeno effect. We observe rapid break-up of bound pairs for weak interactions, and for stronger interactions we see doublon decay rates that are asymmetric when changing from attractive and repulsive interactions, and which strongly depend on the interactions and on-site loss rates. By comparing our experimental data with a theoretical analysis of few-body dynamics, we show that these features originate from a non-trivial combination of dissipative dynamics described by a lattice model beyond a standard Bose-Hubbard Hamiltonian, and the modification of three-atom dynamics on a single site, which is generated alongside strong three-body loss. Our results open new possibilities for investigating bosonic atoms with strong three-body loss features, and allow for the better understanding of the parameter regimes that are required to realize strong effective three-body interactions.
\end{abstract}

\maketitle

Ultracold atomic gases in optical lattices provide a platform for investigating novel many-body coherent and dissipative dynamics \cite{Bloch2008,Lew_book,Bloch:2012aa,Mu_rev,Daley:2014aa}.
In particular, processes such as collisional loss of atoms, which we often seek to avoid, can exhibit features such as the continuous quantum Zeno effect \cite{Syassen1329,Yan:2013aa,Zhu2013,PhysRevA.82.022120},
where a strong dissipative process can prevent coherent dynamics from taking place \cite{PhysRevA.41.2295,PhysRevA.48.132}. This has resulted in proposals to realize effective three-body interactions through three-body loss \cite{Daley2009,PhysRevLett.104.096803,PhysRevB.82.064509,PhysRevB.82.064510}.
However, there are important open questions related to how these dynamics work when the loss rates become comparable to or larger than the energy band gap, rendering a standard Bose-Hubbard (BH) description insufficient. In this work, we explore the interplay between coherent and dissipative dynamics for a gas of bosonic atom pairs, comparing the dependence of the three-body loss on the interaction strength to theoretical models beyond the standard BH Hamiltonian that combine lattice dynamics with a careful treatment of on-site three-body dynamics that renormalizes the coefficients of coherent interactions and loss. Understanding these effects across different regimes provides a path for future studies of exotic quantum phases induced by strong local three-body loss \cite{Schmidt2009,Petrosyan2007,PhysRevB.84.092503,PhysRevA.84.021601,PhysRevA.81.061604,PhysRevLett.106.185302,PhysRevLett.103.240401,PhysRevB.79.020503}.

We make use of the control available  in optical lattice systems and study interacting Cesium atoms in the vicinity of a broad Feshbach resonance \cite{Kraemer2006}, which allows us to tune the $s$-wave scattering length $a_{\rm S}$ and therefore the interactions from weak to strong and from repulsive ($a_{\rm S}>0$) to attractive ($a_{\rm S}<0$). The two-body properties of our system  are well understood \cite{Mark2011,Mark2012,Jurgensen2014} and it is straight-forward to prepare an initial state with on-site pairs (often referred to as doublons) \cite{Winkler2006,Strohmaier2010}. Doublons are in general stable in optical lattice systems because of the lack of dissipative phonon modes that can remove energy on short timescales \cite{Winkler2006,Deuchert2012}. However, by increasing the interaction strength one enters a regime where the energy associated with strong three-body dissipative processes \cite{dincao2018JPB,braaten2006PRep} can easily exceed the band gap in the lattice. This raises questions on how to treat this system with an effective BH model \cite{PhysRevLett.104.090402,Mark2011,Mark2012,Jaksch:2005aa,PhysRevLett.91.090402}, including the question whether off-site loss mechanisms become important in understanding the resulting dynamics, and how effective model parameters would need to be rescaled in that case.  

Below, we investigate the effects of strong three-body losses starting with doubly occupied sites and find unexpected phenomena arising from an off-resonant loss process where a single particle tunnels into a virtual triply occupied site. For strong attractive interactions, we observe an interesting nonlinear dependence of the doublon decay rates on $a_{\rm S}$, and attempt to model this with a BH-type model resulting in values for decay rates that are too large if nearest neighbor losses are included, but too low if they are simply ignored. This indicates that the mixing with higher energy bands is causing a renormalization of both the off-site and the on-site coefficients. When further increasing the attractive interaction strength, we observe a decrease on the decay rate similar to a quantum Zeno type suppression \cite{PhysRevA.48.132}. However, through first-principle calculations of the on-site losses induced through mixing with three-body Efimov states \cite{supmat,goban2018NT,wang2011PRA,efimov1973NPA,dincao2018JPB,braaten2006PRep}, we find that on-site loss rates are too low to induce this type of suppression and we instead show that the decrease of the decay rates is due to an additional renormalization of the terms responsible for the coherent dynamics.

\begin{figure*}[t!]
\includegraphics[width=\textwidth]{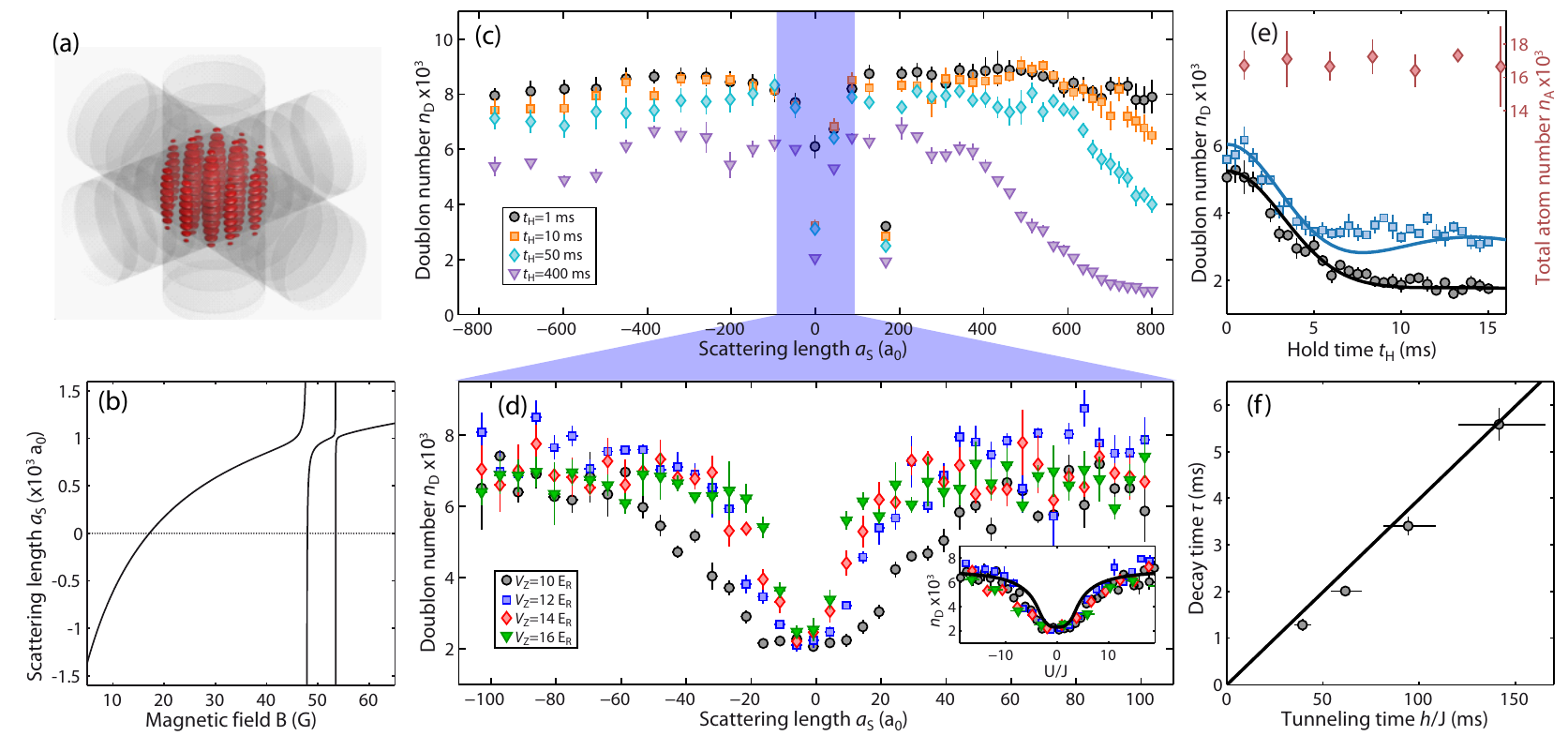}
\caption{(a) Illustration of the lattice 
system. (b) Dependence of $a_{\rm\,S}$ on the magnetic offset field $B$. (c) Number of doublons $n_{\rm\,D}$ as a function of $a_{\rm\,S}$ for $t_{\rm\,H}$ as indicated for $V_{\rm\,x,y,z}\,{=}\,(30,30,10)\,E_{\rm\,R}$. The loss feature near $a_{\rm S}=180$a$_0$ results from a narrow Feshbach resonance. (d) Number of doublons $n_{\rm\,D}$ as a function of $a_{\rm\,S}$ for $t_{\rm\,H}\,{=}\,100\,$ms with $V_{\rm\,z}$ as indicated.
The inset plots the same data as a function of $U/J$, with the solid line showing the numerical result (see text). (e) Number of doublons $n_{\rm\,D}$ as a function of $t_{\rm\,H}$ for $U/J\,{\approx}\,0$ (circles) and $U/J\,{=}\,3.5$ (squares) at $V_{\rm\,z}\,{=}\,14\,E_{\rm R}$ and $V_{\rm x,y}\,{=}\,30\,E_{\rm R}$. The total atom number $n_{\rm\,A}$ (diamonds) for $U/J\,{\approx}\,0$ is also shown as a reference. The solid lines are the corresponding numerical results. (f) Decay time $\tau$, obtained from exponential fits to the decay curves for $U/J\,{\approx}\,0$ for different $V_z$, as a function of the calculated tunneling time $h/J$. The solid line is the estimated decay time $\tau\,{=}\,\hbar/(4J)$. The vertical and horizontal error bars reflect the one-sigma statistical error and a 5\,\%  uncertainty in $V_{\rm\,z}$, respectively. \label{Fig1}}
\end{figure*}

Our experiments start with the production of an essentially pure Bose-Einstein condensate (BEC) of $1.0\times10^{5}$ Cs atoms in the lowest hyperfine state  in a crossed optical trap \cite{Kraemer2004}. To prepare a comparatively dilute system of doublons, we non-adiabatically load the BEC within $2\,$ms into a cubic optical lattice formed by three retro-reflected laser beams with a wavelength of $\lambda\,{=}\,1064.5\,$nm, see Fig.\,\ref{Fig1}(a). Initially, the lattice depths along all three directions are set to $V^0_{\rm\,x,y,z}\,{=}\,30\,E_{\rm\,R}$, where $E_{\rm\,R}\,{=}\,h^2/(2m\lambda^2)$ is the recoil energy with the mass $m$ of the Cs atom. During lattice loading $a_{\rm\,S}$ is set to $220\,$a$_{\rm\,0}$, where a$_{\rm\,0}$ is Bohr's radius. The doublon sample is then purified by Feshbach molecule formation, cleaning and dissociation \cite{PhysRevLett.111.053003}.

Next, we tune $a_{\rm\,S}$ to a value between $-800\,$a$_{\rm\,0}$ and $+800\,$a$_{\rm\,0}$ by ramping the offset magnetic field $B$ with a rate of $\sim\,2.5\,$G/ms to a value between about $8\,$G and $37\,$G, exploiting a broad Feshbach resonance, see Fig.\,\ref{Fig1}(b) \cite{Lange2009}. Note that the tuning range includes a zero crossing for $a_{\rm\,S}$ located at $17.119\,$G. To induce dynamics of the doublons, we quench the lattice depth along the $z$-direction within $1\,$ms to a new value $V_{\rm\,z}$, realizing nearly isolated 1D BH systems. After a variable hold time $t_{\rm\,H}$ we interrupt the dynamics by raising $V_{\rm\,z}$ in $1\,$ms back to its original value. Finally, we determine the remaining number of doublons by employing the association-cleaning-dissociation procedure a second time before detecting the atom number after a short time-of-flight.

The doublon number $n_{\rm\,D}$ shows a strong dependence on $a_{\rm\,S}$ as can be seen in Fig.\,\ref{Fig1}(c) for four different hold times $t_{\rm\,H}$. Clearly, our data exhibits a pronounced asymmetry between strongly attractive and strongly repulsive interactions. We first focus on the region near the zero crossing. A close-up of this region is presented in Fig.\,\ref{Fig1}(d) for four different values of $V_z$ and a hold time $t_{\rm\,H}=100$ ms for which the dynamics has reached a steady state. A pronounced dip for $n_{\rm\,D}$ centered at $a_{\rm\,S}=0$ is evident. In this regime of comparatively weak interactions a standard lowest-band BH description \cite{Jaksch1988,supmat} should be valid. The dip, which is essentially symmetric as a function of $a_{\rm\,S}=0$, results from rapid dissociation of the doublons when the BH interaction energy $|U|$ and the tunneling energy $J$ become comparable, irrespective of the sign of $U$. Indeed, when plotting the data as a function of $U/J$ we observe a collapse onto a single curve, see inset to Fig.\,\ref{Fig1}(d). Note that $n_{\rm\,D}$ remains non-zero at $U/J=0$, which is a consequence of the finite size of the 1D systems, resulting in a non-zero probability to find two atoms on the same lattice site.

We stress that in this regime of comparatively weak interactions, the doublons dissociate but atoms are not lost from the sample. This is evident from Fig.\,\ref{Fig1}(e), which depicts the dissociation dynamics for two values of $U/J$. While the total atom number stays constant, we observe a rapid decay for $n_{\rm\,D}$ with a rate independent of $U/J$ towards the steady-state doublon number shown in Fig.\,\ref{Fig1}(d). Moreover, the initial decay time $\tau$, obtained from an exponential fit to the data, provides a direct measure of $J$ and obeys the linear relation $\tau\,{=}\,\frac{\hbar}{4J}$, see Fig.\,\ref{Fig1}(f).

So far, our observations near $a_{\rm\,S}\,=\,0$ are all captured by the standard lowest-band BH model. To demonstrate this, we numerically calculate the dynamics by exact diagonalization \cite{supmat}. Comparison with the experimental data as shown in Fig.\,\ref{Fig1}(e) gives good agreement. The small deviations probably arise from finite size effects. Also, the steady-state doublon number is well captured, see the inset to Fig.\,\ref{Fig1}(d). 

\begin{figure}
\includegraphics[width=8.5cm]{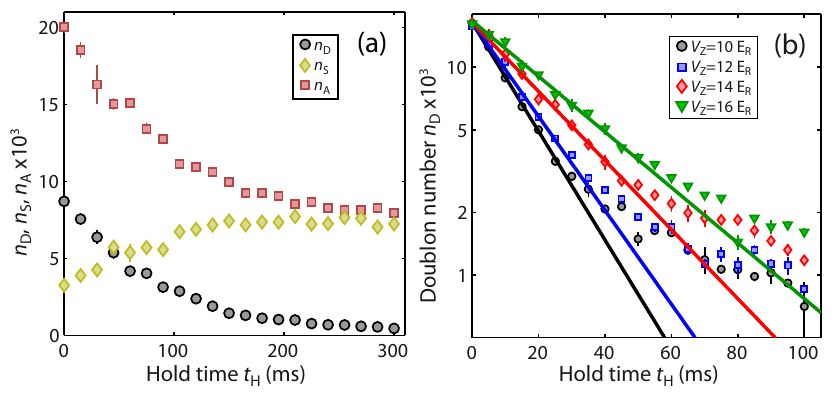}
\caption{(a) Doublon decay for strong interactions: Number of doublons $n_{\rm\,D}$ (circles), singly occupied sites $n_{\rm\,S}$ (diamonds) and total atom number $n_{\rm\,A}$ (squares) as a function of hold time $t_{\rm\,H}$ for a dilute sample for $a_{\rm\,S}\,{=}\,800\,$a$_{\rm\,0}$ and $V_{\rm\,z}\,{=}\,14\,E_{\rm\,R}$. (b) Doublon decay for strong interactions and high density: Number of doublons $n_{\rm\,D}$ as a function of $t_{\rm\,H}$ for $a_{\rm\,S}\,{=}\,800\,$a$_{\rm\,0}$ for $V_{\rm\,z}$ as indicated. The solid lines are fits by an exponential decay to the datapoints with more than 50\% of the initial value, giving decay times $\tau\,{=}\,16.6(0.7), 19.2(1.1), 26.1(0.8)$, and $ 32.5(1.3)\,$ms.\label{Fig2}}
\end{figure}

Let us now turn to the unexplored regime of strong interactions, where the standard BH description no longer applies. As evident from Fig.\,\ref{Fig1}(c), we observe increased loss of doublons with increasing positive $a_{\rm\,S}$, but not so for negative $a_{\rm\,S}$. Figure \ref{Fig2}(a) shows the typical time evolution of  $n_{\rm\,D}$ for strong repulsive interactions, along with the number of singly occupied sites $n_{\rm\,S}$ and the total atom number $n_{\rm\,A}$. In contrast to the case of weak interactions, the loss of double occupancy is caused by a true loss of atoms from the system. We observe that the number of singly occupied sites increases until all doublons are lost. The ratio of lost doublons to lost atoms is $2.1(2)\,:\,3$, the ratio of lost doublons to reappearing atoms at singly occupied sites is $2.1(3)\,:\,1$. This represents strong evidence that the dominant loss process is three-body recombination, where two neighboring doublons exhibit a small probability to decay via a triply occupied site, leaving one singly occupied site behind. Note that this mechanism differs from the elastic decay observed for doublons composed of fermions \cite{Strohmaier2010}.

For a quantitative comparison with a theoretical model, we perform the same experiment in a dense system, i.e. starting from a Mott shell with predominant double occupancy, cleaned from singly occupied sites in the same way as before. This ensures that tunneling dynamics of the doublons themselves do not play any role in the initial dynamics. We estimate the filling fraction to be close to unity for an initial length of about 20 sites for the 1D systems. Figure\,\ref{Fig2}(b) depicts typical decay curves for $n_{\rm{D}}$ from which we deduce initial doublon decay rates $\gamma_{\rm\,D}$ from exponential fits to the data. The decay rates slow down with increased loss as the system gets more dilute. We hence use only the initial parts of the loss curves for the fit.

\begin{figure}
\includegraphics[width=8.5cm]{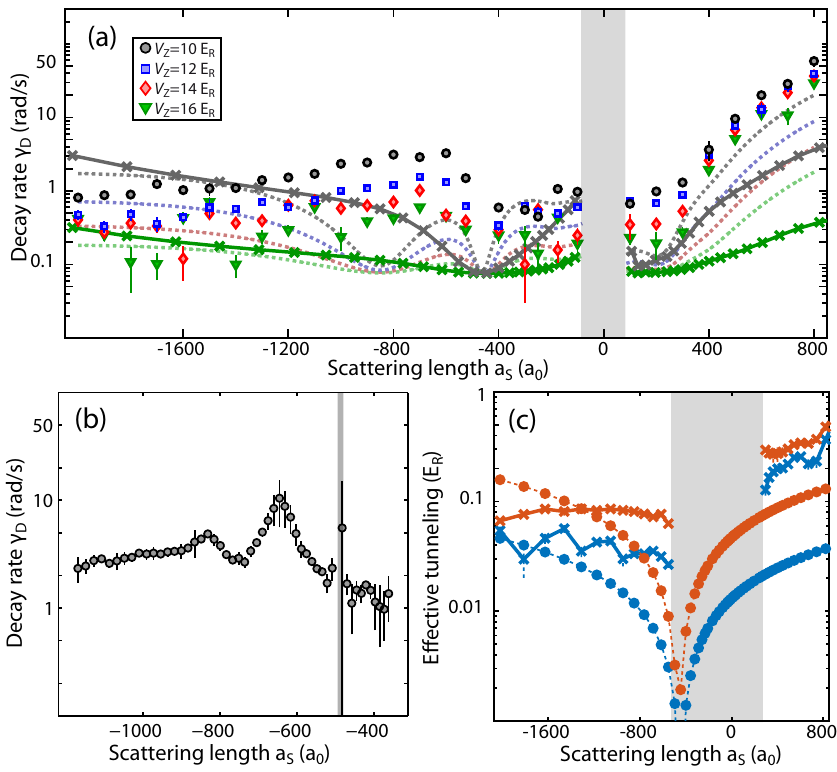}
\caption{Decay rates for strong interactions and high density. (a) Doublon decay rate $\gamma_{\rm\,D}$ starting from the two-atom Mott shell for the full interaction range with $V_{\rm\,z}$ as indicated plotted against a$_{\rm\,S}$. 
Dotted lines (solid lines with crosses) are predictions without (with) modified on-site parameters (see text). Line colors are chosen to match the corresponding experimental data. The shaded area indicates the region of weak interactions where $U\gg J$ is not fulfilled. (b) Fine scan of $\gamma_{\rm\,D}$ for attractive interactions in the range from $-1100\,$a$_{\rm\,0}$ to $-400\,$a$_{\rm\,0}$ with $V_{\rm\,z}\,{=}\,10\,E_{\rm\,R}$. The shaded area indicates the presence of a narrow Feshbach resonance. (c) Renormalized tunneling coefficient within the extended BH model (dotted line with filled circles) and determined from the experimental data (solid line with crosses), for $V_{\rm\,z}\,{=}\,10$ (red) and $V_{\rm\,z}\,{=}\,16$ (blue). The shaded area denotes the region where the perturbative treatment for the latter is not valid.
\label{Fig3}}
\end{figure}

In this way, we measure $\gamma_{\rm\,D}$ for different values of $a_{\rm\,S}$ and $V_{\rm\,z}$. Figure\,\ref{Fig3}(a) shows $\gamma_{\rm\,D}$ for a very broad range of interaction strength, from $-2000\,$a$_{\rm\,0}$ to $+800\,$a$_{\rm\,0}$, revealing the drastic asymmetry between repulsive and attractive interactions that could already be seen in Fig.\,\ref{Fig1}(c). Note that we limit the hold time to $400\,$ms to avoid tunneling dynamics between neighboring 1D systems. This, however leads to comparatively large uncertainties for small decay rates.

We model this behavior by an extended BH model that accounts for on-site and nearest-neighbor two-body interactions as well as dissipative on-site three-body loss \cite{supmat}. Note that we neglect terms corresponding to site-to-site three-body loss. For large values of $|a_{\rm\,S}|$, one might expect to obtain large contributions from off-site loss processes. However, this picture only holds for sufficiently small loss rates compared to the band separation energy when the lowest-band Wannier functions provide a good local basis. In the presence of large loss rates, as is the case for large $|a_{\rm\,S}|$, the Wannier states mix with states from higher bands and we find that the initial overlaps of nearest-neighboring wavefunctions lead to a very rapid initial decay, although only resulting in a very small loss of atoms. This happens on timescales much faster than the experimental measurements, resulting in renormalized states that are more localized on each lattice site. Atom loss is then driven by on-site decay only, combined with non-resonant tunneling processes, including density-assisted tunneling from nearest neighbor interactions.

To understand the importance of competing on-site and tunneling processes in the dynamics of particle loss, we perform calculations in which we evolve the initial doubly-occupied Mott state using a Lindblad master equation. We incorporate three-body loss, and also single-particle loss terms that account for residual heating induced by the lattice light and/or background gas collisions \cite{Mark2012}. We extract the decay rate $\gamma_{\rm\,D}$ from an exponential fit to the computed doublon density. In order to identify the roles of different processes, we use two complementary methods to estimate coefficients for our model.

In the first approach, we calculate lattice coefficients using single-particle Wannier functions associated with the lowest Bloch band. On-site loss is modeled as a zero-range three-body recombination with a rate coefficient $L_3$, with parameters for Cs fitted to the experimental data from Ref.~\cite{Kraemer2006}.  As we will see below, the resulting model predicts decay rates at the same order of magnitude as the experimental measurements, capturing  qualitative aspects well for positive scattering lengths. However, it fails to predict the finer details for negative scattering lengths due to the presence of a large three-body resonance, for which we clearly  must go beyond a standard BH model description.

In our second approach, we determine more accurate on-site parameters by taking the lowest on-site three-body energy level obtained by using the hyperspherical adiabatic representation \cite{dincao2018JPB,wang2011PRA}. This allows us to calculate, from first principles, the on-site three-body losses from and energy shifts to the lowest band due to three-body interactions mixing with higher bands \cite{goban2018NT}. For on-site two-body interactions we also take into account mixing with higher bands using the result presented in Ref.~\cite{BuRes} for a single harmonic trap. We find that by incorporating these corrections from strong interactions the main features near the resonance are better approximated. However, we do not modify the nearest-neighbor tunneling and interaction terms, and find that these must also play an important role away from the resonance.

In Fig.\,\ref{Fig3}(a) we compare $\gamma_{\rm\,D}$ obtained from the two approaches with the experimental data. We find that for repulsive interactions the decay rates increase with increasing $a_{\rm\,S}$, in accordance with the predictions. We also find that the first approach (dotted lines) predicts a minimum in the decay rate around $a_{\rm{S}}=-800a_0$, which coincides with the strong Efimov resonance reported in $L_3$ \cite{Kraemer2006}, and which is not observed in our lattice experiment.
Our second theoretical approach (solid lines with symbols), incorporating the effects of the lattice potential on the loss rates, predicts a suppression of the $L_3$ resonance and we obtain values for $\gamma_{\rm\,D}$ that qualitatively match the experiment \cite{supmat}. This peak suppression is due to the interplay between the Efimov resonance, the mixing with higher bands and the lattice potential, which creates a spread of the particles in momentum space. Indeed, this spread is of the same magnitude as the temperature-induced energy spread that was found to suppress the Efimov resonance in previous measurements \cite{Kraemer2006,dincao2004PRL}. In addition, effects of mixing with higher excited states introduce further modifications to the on-site, but also the off-site, loss coefficients producing the double peak feature shown in Fig~\ref{Fig3}(b).  

Furthermore, including corrections from strong interactions (solid lines with symbols in Fig.~\ref{Fig3}(a)) predicts that, for increasingly large attractive interactions, the decay rates also increase, in sharp contrast to the experimental data. We initially interpreted the trend in the measurements to be due to a quantum-Zeno-type suppression of the losses \cite{PhysRevA.48.132}, but the values of the renormalized coefficients in our calculation predict on-site loss rate values that are too low for this to occur \cite{supmat}. 

Instead, we believe that this trend can be accounted for through the incorporation of two effects not currently taken into account in our model: mixing between ground and excited on-site three-body states \cite{supmat}, and a further renormalization of the effective tunneling elements \cite{Juergensen2014}. Our experiment enables us to estimate what the values of the rescaled tunneling coefficients, $\tilde{J}$, would need to be to produce the measured decay rates, accounting for interaction-modified tunneling coefficients \cite{supmat}. These are shown in Fig.\,\ref{Fig3}(c), and predict a saturation of the magnitudes of the rescaled tunneling rate for negative $a_{\rm\,S}$, an effect that is more pronounced for shallow lattices. Additionally, for large positive $a_{\rm\,S}$ the rescaled tunneling rates are significantly enhanced.

In summary, we have investigated the dynamics of bosonic doublons for a broad range of attractive and repulsive interactions. We have found that for strong two-body interactions (attractive and repulsive) the dynamics are dominated by off-resonant decay processes induced by on-site three-body loss, with a large renormalization of the coefficients in both the dissipative and coherent processes, which can be understood by solving the three-body problem on-site. For particularly strong interactions, we enter a further regime, where strong interactions lead to further corrections, which can be attributed to a combination of modified effective tunneling rates, and further mixing of on-site three-body eigenstates.

In this way, our experiments and theoretical analysis both answer existing and (for stronger interactions) open new questions on the most accurate means to model dissipation for atoms in optical lattices. This is important for high-efficiency molecular formation \cite{Thalhammer2006}, but also opens routes to further explore physics beyond the Bose-Hubbard model, and as well as to identify regimes of true quantum Zeno effect suppression of three-body occupation. This opens routes towards a range of exotic many-body states driven by three-body projections.

\begin{acknowledgments}
We thank Suzanne McEndoo for important discussions during the early stages of this work. The Innsbruck team gratefully acknowledges funding by the European Research Council (ERC) under project no. 278417, under project no. 789017 and by the Austrian Science fund (FWF) via the FWF-ANR joint project with project no. I2922-N36. This work was supported in part by the European Union's Horizon 2020 research and innovation program under grant agreement No. 817482 PASQuanS. S.F. acknowledges the financial support of the Carnegie trust. J.P.D. acknowledges support from National Aeronautics and Space Administration (NASA/JPL 1502690). F.M. acknowledges support by the Carl Zeiss Foundation via IQST and is indebted to the Baden-W\"urttemberg-Stiftung for the financial support by the Eliteprogramm for Postdocs.
\end{acknowledgments}

\appendix
\renewcommand\thefigure{\thesection S\arabic{figure}}   
\setcounter{figure}{0}   
\section{Supplementary Material}

\section*{Exact diagonalization standard BH Model}\label{standardmodel}

We perform exact diagonalization of small 1D systems of typically $6$ atoms on $11$ lattice sites with hard wall boundary conditions. To mimic the dilute sample of randomly distributed doublons, we compute the time evolution of each possible initial doublon state and average the time evolution of the doublon probability.

\section*{Extended BH Model}\label{model}

The 1D extended BH model that we use consists of on-site and nearest-neighbor two-body interactions and on-site three-body loss. In principle there can also be three-body recombination effects between atoms in nearest neighboring Wannier states. However, we obtain much better agreement with the experiment for negative values for $a_{\rm\,S}$ if we neglect these contributions. This makes physical sense in the limit where the three-body loss rate dominates the dynamics, as any small initial overlap between neighboring states would be very rapidly lost, on timescales much shorter than the experimentally measured dynamics, resulting in more localized on-site wavefunctions. Explicitly, the Hamiltonian is given by ($\hbar=1$),
\begin{equation}\label{Ham}
\begin{split}
H &= -J \sum_{\langle i,j \rangle} b_i^{\dagger} b_j + \frac{U}{2} \sum_i b_i^{\dagger 2} b_i^2 \\
&+ \tilde{U} \sum_i [ b_i^{\dagger}b_i^{\dagger}b_i b_{i+1} + b_i^{\dagger}b_{i+1}^{\dagger} b_{i+1} b_{i+1} + h.c. ] \\
&- i\frac{\gamma_3}{12} \sum_i b_i^{\dagger 3} b_i^3 \, .
\end{split}
\end{equation}
The coefficients, calculated through the Wannier functions associated to the lowest Bloch band centered at position $z_i$ in the longitudinal direction, $w(z-z_i)$, and $\vec{r}_i$ in the radial direction, $w_{\bot}(\vec{r}-\vec{r}_i)$, are given by
\begin{equation} \label{Coeffs_Wann}
\begin{split}
U &= \frac{4\pi\hbar^2 a_s}{m} \int d\vec{r} |w_{\bot}(\vec{r} - \vec{r}_i)|^4 \int dz |w(z - z_i)|^4, \\
\tilde{U} &= \frac{4\pi\hbar^2 a_s}{m} \int d\vec{r} |w_{\bot}(\vec{r} - \vec{r}_i)|^4, \\
& \times \int dz |w(z - z_i)|^{2} w(z - z_i)^* w(z - z_{i+1}), \\
\gamma_3 &= \Gamma_3 \int d\vec{r} |w_{\bot}(\vec{r} - \vec{r}_i)|^6 \int dz |w(z - z_i)|^6 \, . \\
\end{split}
\end{equation}
As before, $a_s$ denotes the scattering length, $J$ is the single particle nearest-neighbor tunneling amplitude, and $\Gamma_3 = 2 L_3$, where $L_3$ is the experimentally measured three-body loss parameter \cite{Kraemer2006}. The approximation of restricting the dynamics to 1D is valid as long as the radial trapping frequency is much larger than the longitudinal trapping frequency, $\omega_r \gg \omega_z$. Note that here, we have taken into account three-body Efimov resonances by incorporating the short distance cut-off term, which has to be introduced to regularize the zero-range three-body pseudo-potential \cite{EFIMOV1970563,PhysRevLett.82.463}, as a prefactor for a three-body delta-function contact interaction \cite{Daley2009,Naidon_2017}. 

In order to extract the decay rates from this model, we consider four Cs atoms in two lattice sites and we begin with the initial state of two atoms on each site. We then apply a Lindblad master equation,
\begin{equation}
\frac{d}{dt} \rho = -i[H_{\rm eff},\rho] + 2L \rho(t) L^{\dagger} + 2L_s \rho(t) L^{\dagger}_s,
\end{equation}
where $L = \sqrt{\gamma_3/12} b_i^3$. We have included a single-particle loss rate, $H_{\rm eff} = H + H_s$, where
\begin{equation}
H_{s} = -i\tilde{S} \sum_i b^{\dagger}_i b_i,
\end{equation}
and $L_s = \sqrt{\tilde{S}}b_i$. For the loss rate, we use a value $\tilde{S} = 0.006 \, \rm{Hz}$, which was fine tuned so that this model would reproduce the lowest experimentally measured value. We then calculate the time-dependent expectation value of the total doublon number in the system, and fit an exponential to this function, $\langle \hat{N}(t) \rangle/N(0) = \exp(-\alpha t)$, to extract the decay rate $\alpha$.

Using the master equation we compare two cases. One case where the model coefficients are given by Eq.~\ref{Coeffs_Wann} with the free space $L_3$ parameter \cite{Kraemer2006}, and second case where we include corrections to the on-site lattice parameters due to couplings to higher energy states induced by two- and three-body interactions. For two-body interactions we make use of the analytical results presented in Ref.~\cite{BuRes}, and we explain how calculations of the three-body states lead us to corrections for the on-site three-body interaction and also give rise to an effective loss rate in the next section. Note that both calculations approximate each lattice site as an isotropic harmonic oscillator: the value of the energy spacing is given by the geometric mean of the three energies of the anisotropic experiment.

If we set the single particle loss rate, $\tilde{S} = 0$, we can calculate the decay rate through second-order perturbation theory in $J \ll U$,
\begin{equation}\label{Pert}
\Delta E^{(2)} = -2\frac{|3\sqrt{6} \tilde{U} - \sqrt{6} J|^2}{E_{3B}(a_s) - 2 U_{2B}},
\end{equation}
where $E_{3B}(a_s)$ and $U_{2B}$ are the energy shifts due to three and two interacting atoms, respectively. The decay rate is then found through 
\begin{equation}
P(t) = \langle \phi | e^{i (\Delta E^{(2)*} - \Delta E^{(2)})t} | \phi \rangle = e^{-i 2 {\rm Im}(\Delta E^{(2)})t},
\end{equation}
giving us an expression for the decay rate, $\alpha = 2 {\rm Im}(\Delta E^{(2)})$.

We then use this expression to find the rescaled value for $\tilde{J} = |3\sqrt{6} \tilde{U} - \sqrt{6} J|$, such that the model (with rescaled on-site coefficients) gives rise to the experimentally observed decay rates. Note that we only perform this fit for larger values of $|a_s|$ (where the perturbation theory is most valid) because for lower values the calculated decay fits the experimental measurements quite well and is dominated by the single-particle losses.

\section*{On-site three-body energy levels and decay}

Our three-body calculations for Cs atoms are performed using the the adiabatic hyperspherical 
representation \cite{dincao2018JPB,wang2011PRA}, where the interatomic interactions are given by a two-channel
model similar to the one used in Ref. \cite{wang2014NP}, which properly describes the $-11$-G Feshbach resonance for the $F=3$, $m_F=3$ hyperfine state of Cs \cite{chin2010rmp}. Our methodology, developed in Ref.~\cite{chapurin2019PRL}, incorporates the use of Feshbach projectors \cite{Jonsell_2004} to represent the fully symmetric three-body spin states as well as the relevant decay channels responsible for atomic losses.
In the hyperspherical representation the hyperradius $R$ determines the overall size of the system, while all other degrees of freedom are represented by a set of hyperangles $\Omega$. Within this frame work, the three-body adiabatic potentials $U$ and channel functions $\Phi$ are determined from the solutions of the hyperangular adiabatic equation:
\begin{align}
&\left[\frac{\Lambda^2(\Omega)+15/4}{2\mu R^2}\hbar^2+\varepsilon_{\alpha}\right]\Phi_{\alpha}(R;\Omega)\nonumber\\
&+\sum_{\beta}\sum_{i<j}V_{\alpha\beta}(r_{ij})\Phi_{\beta}(R;\Omega)=U(R)\Phi_{\alpha}(R;\Omega),
\label{AngEq}
\end{align}
which contains the hyperangular part of the kinetic energy, expressed through the grand-angular momentum operator $\Lambda^2$ and the three-body reduced mass $\mu=m/\sqrt{3}$. In the above equation, our two-channel threshold energies and model potential are given, respectively, by
\begin{align}
\varepsilon=\left(\begin{array}{c} \varepsilon_{\rm bg} \\ \varepsilon_{\rm res} \end{array}\right)~~\mbox{and}~~~
V(r)=\left(\begin{array}{cc} V_{\rm bg}(r) & V_c(r)\\ V_c(r) & V_{\rm res}(r) \end{array}\right),
\end{align}
where the background and resonant interactions are $V_{\rm bg}(r)=-C_6/r^6(1-\lambda_{\rm bg}^6/r^6)$
and $V_{\rm res}(r)=-C_6/r^6(1-\lambda_{\rm res}^6/r^6)$, with $C_6=6890.4768$ being the van der Waals dispersion 
coefficient for Cs \cite{berninger2013PRA}, and interchannel coupling $V_c(r)=\alpha_c\exp[-(r-\beta_c)^2/2\gamma_c^2]$.
Assuming that $\varepsilon_{\rm bg}=0$, we choose the parameters $\lambda_{\rm bg}$, $\lambda_{\rm res}$, $\alpha_c$,
$\beta_c$, $\gamma_c$ and $\varepsilon_{\rm res}$ in order to properly reproduce the magnetic-field dependence of the 
$-11$-G Feshbach resonance. Our calculations are performed assuming $V_{\rm bg}$ 
and $V_{\rm res}$ each containing four $s$-wave molecular states. The values for the energy and corresponding decay are then obtained by solving the hyperradial Schr\"odinger equation \cite{wang2011PRA} in the presence of a harmonic confinement and leaving the adiabatic channels associated to the diatomic states open, thus allowing the state to decay \cite{goban2018NT}. 

\begin{figure}
\includegraphics[width=8.5cm]{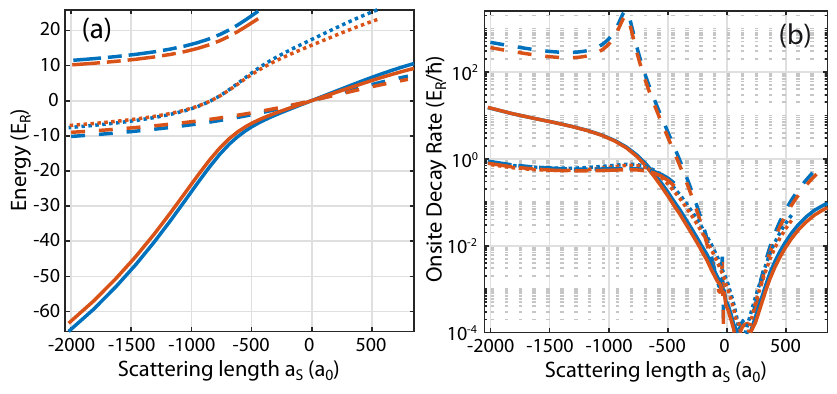}
\caption{On-site three-body energies (left) and on-site loss rates (right) calculated using the the adiabatic hyperspherical representation \cite{dincao2018JPB,wang2011PRA}, for $V_z=16 E_R$ (blue) and $V_z=10 E_R$ (red). We include the first, second and third three-body state (solid, dotted, dashed-dotted). We also include the (two-body) interaction energy of the initial state of two atoms on each lattice site (left, dashed), calculated using Ref.~\cite{BuRes}. And we include the on-site loss rate calculated from the BH model using the free-space $L_3$ parameter \cite{Kraemer2006} (right, dashed). \label{Supp_Fig1}}
\end{figure}

The explicit energy and loss values for the lowest and first two excited three-body states are given in Fig.~\ref{Supp_Fig1}. We also include the interaction energy of the initial state in the experiment: two atoms on each lattice site. This allows us to compare the energy shift (see Eq.~\ref{Pert}) that the system is required to overcome for each three-body state when the system tunnels from two doubly occupied sites into the triply occupied site, which is the non-resonant process responsible for the atomic loss in the experiment. We can see that for positive and small negative values of $a_{\rm\,S}$ the three-body state closest to being resonant is the lowest, however, for large and negative values of $a_{\rm\,S}$, the most resonant state is the first excited one. This may indicate that this is the most likely state that is populated when the third particle tunnels into the site, which leads us to believe that the remaining discrepancies between our theory predictions and the experimental data presented in the main text could be due to couplings into these excited states. 

However, the energy resonance condition is not the whole picture, there is still the question of the rescaling of the effective tunneling rates, which complicates this conclusion. And so it is likely that the correct on-site three-body state to use is a non-trivial superposition between the lowest and the first couple of excited three-body eigenstates. We leave the analysis of the coefficients in this superposition as a future objective.

In Fig.~\ref{Supp_Fig1}(b), we can see that for large negative scattering lengths the renormalized values for the onsite decay rates are two orders of magnitude smaller than those calculated with the free space $L_3$ parameter (see $\gamma_3$ in Equ.~\ref{Coeffs_Wann}). This indicates that the onsite losses are too low to induce the quantum Zeno suppression usually expected in this regime. 

\section*{Future Theoretical Analysis}
In this section we summarize some of the important approximations that go into deriving our models. The purpose of this is to highlight the additional features that should be incorporated for future theoretical studies into the phenomena presented in the main text.

\begin{itemize}

\item \textit{On-site coefficient calculation}: This calculation assumes an isotropic harmonic oscillator for a single lattice site, where we have taken the geometrical mean of the energy levels in the three directions in the experimental set-up. The interplay between the anisotropy and the free-space Efimov resonance can allow for additional features on the observed decay rates as displayed in Fig.\,\ref{Fig3}(c).

\item \textit{Mixing of on-site three-body eigenstates}: As shown in Fig.~\ref{Supp_Fig1}, particularly for large and negative values of $a_{\rm\,S}$, the energy of the initial state (dashed line) is very close to being resonant with the first excited three-body eigenstate, indicating that in this regime there is a strong possibility that these states will also be important and result in a further rescaling of the on-site loss rates. In the main text, we have attempted to predict the values of the renormalized tunneling terms that couple the initial state into a triply occupied site, but for this calculation we ignored the excited three-body eigenstates. One would then need to calculate the individual matrix elements describing the coupling of the initial state into each of the three-body eigenstates separately in order to better predict the overall decay rates.

\item \textit{Restricting the dynamics to 1D}: The strong coupling with states in higher bands may be enhancing the single-particle tunneling in the $z$-direction and potentially increasing tunneling in the other two directions. Such effects could be reducing the validity of our 1D approximation.   

\end{itemize}

\end{document}